\documentclass[aps, prb, twocolumn, superscriptaddress, longbibliography, 10pt, floatfix]{revtex4-2}
\usepackage{silence}
\usepackage{amsmath, amssymb}

\usepackage{physics}
\usepackage{xcolor}
\usepackage{bm}
\usepackage{pdfpages}
\usepackage[colorlinks=true, urlcolor=blue, linkcolor=blue, citecolor=blue]{hyperref}
\newcommand{\Abk}{A(k_y l_y)}
\newcommand{\EF}{\varepsilon_\mathrm{F}}
\newcommand{\matname}{Nb$_{2n+1}$Si$_n$Te$_{4n+2}$ }

\makeatletter
\AtBeginDocument{\let\LS@rot\@undefined}
\makeatother

\begin{document}

\title{Optical conductivity of topological semimetal Nb$_{2n+1}$Si$_n$Te$_{4n+2}$}
\author{Seongjin Ahn}
\affiliation{Department of Physics, Chungbuk National University, Cheongju 28644, Republic of Korea}

\begin{abstract}
We study the linear optical conductivity of the \matname family of layered van der Waals materials, which has recently gained considerable attention owing to its dimensionality-tunable electronic structure with a quasi-one-dimensional nodal-line state. At zero temperature, we analytically show that the Drude weight exhibits strong anisotropy: along the transverse direction it is finite at charge neutrality, whereas in the nodal-line direction it vanishes quadratically with Fermi energy. On the other hand, the interband optical conductivity exhibits the same linear frequency dependence along both the longitudinal and transverse directions, with only a direction-dependent slope in the low-frequency regime. We further analyze the leading finite-temperature corrections to the intraband and interband optical conductivities, showing that the zero-temperature results remain valid up to experimentally relevant temperatures.
\end{abstract}

\maketitle
\section{Introduction}
The material dimensionality is of central importance in condensed matter physics as it governs the physical properties of materials. Reducing the dimensionality in particular can have strong consequences. For example, one-dimensional (1D) transport is governed by the universal quantum of conductance $e^2/h$, where Planck's constant $h$ appears explicitly, indicating that 1D transport is an intrinsically quantum phenomenon with no classical analog \cite{vanWees1988QuantizedConductance,Voit1995OneDimensionalFermiLiquids}. The universal quantum of conductance also characterizes the Drude weight for 1D free fermions, $D=e^2v_\mathrm{F}/h$, where $v_\mathrm{F}$ is the Fermi velocity \cite{Voit1995OneDimensionalFermiLiquids}.
This is in sharp contrast to two-dimensional (2D) and three-dimensional (3D) systems, where in the semiclassical regime the Drude weight, $D = \pi \rho e^2/m$, and the optical conductivity, $\sigma = \rho e^2\tau/m$, both depend on material parameters such as carrier density $\rho$, effective mass $m$, and scattering time $\tau$, with no explicit dependence on Planck's constant $h$ \cite{Pronin2021SurveyOpticalConductivity}.

The family of composition-tunable compounds \matname ($n=1,2,\ldots,\infty$) has recently attracted considerable attention, both theoretically and experimentally, as a platform hosting dimensionality-tunable electronic systems \cite{Zhu2020TunableUnidirectional1DSystem,yangDirectionalMasslessDirac2020,wangOnedimensionalMetalEmbedded2021,zhangObservationDimensioncrossoverTunable2022}.
Structurally, the \matname family may be viewed as NbTe$_2$ metallic chains embedded in a 2D semiconductor with the integer $n$ controlling the separation between neighboring chains (and thus the interchain coupling strength). For $n=1$, i.e., Nb$_3$SiTe$_6$, the system hosts a 2D nodal-line state protected by the nonsymmorphic glide-mirror symmetry \cite{liNonsymmorphicsymmetryprotectedHourglassDirac2018,Liu2022CoexistenceHourglassNodalLine,Liu2023DiracNodalLineNb3SiTe6}. With increasing $n$, the nodal-line state undergoes a dimensional crossover from a 2D state to a 1D-like state as the coupling between neighboring metallic chains is reduced \cite{Zhu2020TunableUnidirectional1DSystem,zhangObservationDimensioncrossoverTunable2022,caoPlasmonsTwodimensionalNonsymmorphic2023,zhaoBerryCurvatureDipole2023}.
It has been theoretically revealed that the quasi-1D nodal-line states give rise to unusual physical properties, including strongly anisotropic plasmons whose intraband frequency is independent of carrier density normal to the nodal line~ \cite{caoPlasmonsTwodimensionalNonsymmorphic2023} and a pronounced Berry curvature dipole and nonlinear Hall effect~ \cite{zhaoBerryCurvatureDipole2023}.

Despite this broad experimental and theoretical interest, the optical conductivity of \matname has remained poorly understood. Optical conductivity encodes the response to an oscillating electromagnetic field, and is thus a powerful tool for probing the electronic properties of materials. In particular, it has been widely used to investigate topological semimetals \cite{Pronin2021SurveyOpticalConductivity,EbadAllah2023OpticalConductivityNb3SiTe6,caoCaoOpticalSignatureFlat2025,3xs5-km1v}. For example, in 3D nodal-line semimetals, the optical conductivity at low frequencies approaches a constant or exhibits a simple power-law behavior depending on the nodal-line geometry  \cite{ahnElectrodynamicsFermiCyclides2017,Barati2017OpticalConductivityNodalLine,mukherjeeTransportOpticsNode2017,Jeon2023OpticalTransitionsSingleNodalRing,Shao2019OpticalSignaturesNbAs2}, while for multi-Weyl nodes the optical conductivity obeys a power-law relation directly derived from the winding number \cite{Ahn2017OpticalConductivityMultiWeyl}. Such distinct frequency dependences arise from the quantum state structure and serve as a probing tool for the electronic structure of topological semimetals.\cite{Shao2019OpticalSignaturesNbAs2,Pronin2021SurveyOpticalConductivity,Jeon2023OpticalTransitionsSingleNodalRing}.
For \matname family materials, the optical conductivity for Nb$_3$SiTe$_6$ ($n=1$ member) has been reported to exhibit strong peaks near 0.15 eV and 0.28 eV due to van Hove singularities \cite{EbadAllah2023OpticalConductivityNb3SiTe6}, along with a peak near 1.2 eV arising from remote flat-band transitions  \cite{caoCaoOpticalSignatureFlat2025}. However, a comprehensive understanding of the optical response for the whole \matname family remains lacking, including the Drude weight behavior and the power-law dependence of the low-frequency optical conductivity.

In this work, we calculate the optical conductivity of \matname using the Kubo formula. We use the Dirac Su-Schrieffer-Heeger (SSH) model that captures the low-energy band dispersion formed by an array of metallic $\mathrm{NbTe_2}$ chains \cite{zhangObservationDimensioncrossoverTunable2022}. This model has been employed in previous studies of \matname, and has captured several exotic phenomena, including unidirectional transport, dimensional crossover behavior, and plasmons with extreme anisotropy \cite{Zhu2020TunableUnidirectional1DSystem,zhangObservationDimensioncrossoverTunable2022,yangDirectionalMasslessDirac2020, gaoIntrinsicHyperbolicityTwodimensional2025}. Our results reveal that the quasi-1D nodal line produces a set of low-energy optical signatures distinct from those of other nodal-line semimetals. We show analytically that the Drude weight along the transverse direction is finite at charge neutrality, inheriting the 1D Dirac Drude weight, while the longitudinal Drude weight vanishes quadratically with Fermi energy, as in an ordinary metal. In contrast to this strongly anisotropic Drude response, interband optical conductivities along both directions grow linearly with frequency at low frequencies despite the anisotropic quasi-1D nodal-line electronic structure. We further derive the leading-order temperature corrections and verify that our zero-temperature results remain valid at experimentally relevant temperatures. Our results offer qualitative and quantitative guidance for optical experiments on the \matname family nonsymmorphic nodal-line materials.

The paper is organized as follows. In Sec.~\ref{sec:theory}, we introduce the Dirac SSH model and the noninteracting Kubo formula used throughout the paper. In Sec.~\ref{sec:zeroTopcd}, we present an analytical and numerical analysis of the zero-temperature optical conductivity arising from the intraband and interband transitions. Section~\ref{sec:finiteTopcd} examines finite-temperature effects, showing that our key findings remain valid up to experimentally relevant temperatures. Section~\ref{sec:conclusion} summarizes our results.

\begin{figure}
    \centering
    \includegraphics[width=\linewidth]{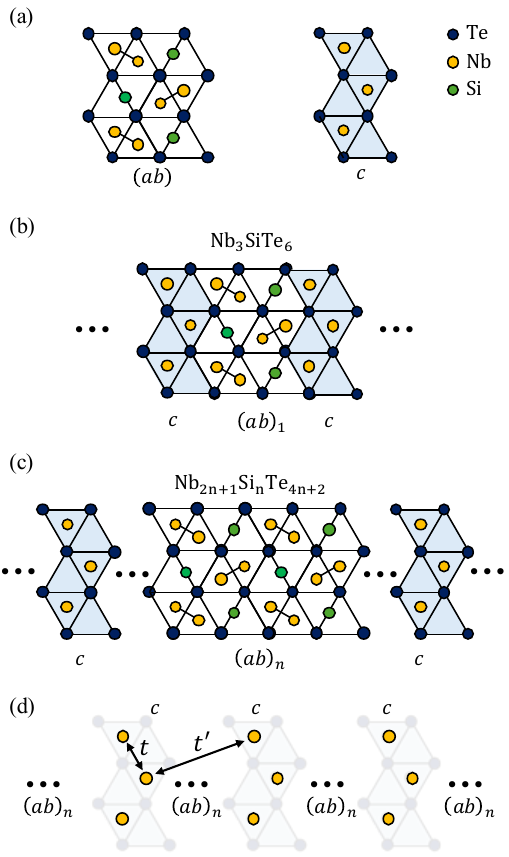}
\caption{ 
Coupled-chain structure of \matname.
(a) Basic building blocks denoted by $(ab)$ and $c$ chains, with blue, yellow, and green circles representing Te, Nb, and Si atoms, respectively.
(b) Structure of the $n=1$ member, i.e., Nb$_3$SiTe$_6$, where periodic $c$ chains are separated by one $(ab)$ block.
(c) General structure of \matname, where neighboring periodic $c$ chains are separated by an $(ab)^n$ spacer.
(d) Rectangular lattice of Nb atoms in the $c$ chains, which corresponds to the Dirac SSH model that describes the low-energy electronic structure near the Fermi level. The parameter $t$ denotes the intrachain hopping along a $c$ chain, while $t'$ denotes the interchain hopping between neighboring $c$ chains through the $(ab)_n$ spacer. Increasing $n$ suppresses $t'$ and drives the system toward the quasi-one-dimensional limit.
} \label{fig:structure}
\end{figure}

\section{Theoretical Formalism} \label{sec:theory}
\subsection{Dirac SSH model}
\begin{figure}
    \centering
    \includegraphics[width=\linewidth]{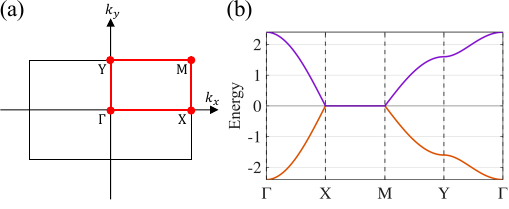}
    \caption{(a) 2D square Brillouin zone showing the high-symmetry points (red dots) and the band-structure path $\Gamma$--X--M--Y--$\Gamma$ (red lines). (b) Energy band structure of the Dirac SSH model with energy measured in units of $t$. We set $t'=0.2t$ for the band calculation.}
    \label{fig:energy_bands}
\end{figure}

In this section, we briefly introduce the Dirac SSH model that captures the low-energy bands forming the nodal line \cite{Zhu2020TunableUnidirectional1DSystem, zhangObservationDimensioncrossoverTunable2022,caoPlasmonsTwodimensionalNonsymmorphic2023}. Figure~\ref{fig:structure} shows the structure of \matname. The \matname family consists of an array of quasi-1D $\mathrm{NbTe_2}$ metallic chains (c-chains in Fig.~\ref{fig:structure}) embedded in a 2D semiconductor [(ab)$^n$-spacers in Fig.~\ref{fig:structure}], with the integer $n$ controlling the separation between neighboring chains. An isolated $\mathrm{NbTe_2}$ chain can be described by an SSH-like model with two sites in each unit cell. Unlike the conventional SSH model, however, the nonsymmorphic glide-mirror symmetry enforces equal nearest-neighbor hoppings along the chain, thus forbidding dimerization and keeping the spectrum gapless. Including the interchain coupling between neighboring chains then yields the following Dirac SSH Hamiltonian for the low-energy bands:
\begin{equation}
\mathcal{H}(\bm{k})
  = \begin{pmatrix} 0 & \phi(\bm{k}) \\ \phi^{*}(\bm{k}) & 0 \end{pmatrix},
\label{eq:H0}
\end{equation}
where $\phi(\bm{k}) = \bigl(t + t'e^{-ik_y l_y}\bigr)\bigl(1 + e^{-ik_x l_x}\bigr)$, $l_x$ and $l_y$ are the lattice constants and $t$($t'$) denotes the intrachain (interchain) hopping in the effective coupled-chain picture [see Fig.~\ref{fig:structure}(d)]. The energy dispersions are given by $\varepsilon_{\pm}(\bm{k})=\pm|\phi(\bm k)|$ with $|\phi(\bm k)|^2 = 4\cos^2\!\frac{k_x l_x}{2}\,\Abk^2$
where $\Abk \equiv \sqrt{t^2+t^{\prime 2}+2tt'\cos k_y l_y}$. 
Note that the nodal line is formed along the Brillouin-zone edge $k_x l_x = \pi$, as shown in Fig.~\ref{fig:energy_bands}. By expanding the energy dispersion around a point on the nodal line [$\bm k_0 = (\pi/l_x, k_y)$] with a small deviation $\delta k_x$ along the $x$ axis, we obtain $\varepsilon_{\pm, \bm k} \approx \pm\hbar v_\mathrm{F}(k_y) \delta k_x$, where 
\begin{equation} \label{eq:vF_qy}
    v_\mathrm{F}(k_y) = l_x \Abk/\hbar
\end{equation} 
is the $k_y$-dependent(and $k_x$ independent) Fermi velocity that describes low-energy dynamics near the nodal line.
Note that this form of dispersion clearly shows the quasi-1D nature of the low-energy physics: for each fixed $k_y$ the low-energy dispersion reduces to a 1D Dirac dispersion along the $x$ direction, while $k_y$ parametrizes the family of chains.

Throughout the work, we set $t'/t=0.2$, which is based on recent ARPES experimental results where the hopping parameters were reported as $t=0.177$ eV and $t'=0.05$ eV for the $n=2$ member \cite{zhangObservationDimensioncrossoverTunable2022}. The ratio $t'/t$ decreases further for members with $n>3$ because increasing $n$ increases the interchain distance and thus reduces the interchain hopping $t'$ (see Fig.~\ref{fig:structure}). As will be shown in the subsequent sections, the optical conductivity shows a quasi-one-dimensional behavior in the small-$t'$ limit because the system approaches a set of weakly coupled one-dimensional chains with $t'\to0$.

\subsection{Kubo formula}
In this work, we limit our scope to the linear optical response. The optical conductivity can be obtained from the Kubo formula \cite{mahanManyparticlePhysics2000}:
\begin{equation}\label{eq:KuboFormula}
\begin{aligned}
    \sigma_{ij}(\omega)= & -g\frac{ie^{2}}{\hbar}\sum_{s, s'}\int_{BZ} \frac{d^{2}k}{(2\pi)^{2}}\frac{f_{s, \bm{k}}-f_{s', \bm{k}}}{\varepsilon_{s, \bm{k}}-\varepsilon_{s', \bm{k}}}\\
    &\times\frac{M_{i}^{ss'}(\bm{k}) M_{j}^{s's}(\bm{k})}{\hbar \omega+\varepsilon_{s, \bm{k}}-\varepsilon_{s', \bm{k}}+i \Gamma}
\end{aligned}
\end{equation}
where $g=2$ is the spin degeneracy, $i, j=x, y$ and $s, s'=\pm$ represent band indices. Here, $f_{s, \bm{k}}\equiv f(\varepsilon_{s,\bm{k}};\mu,T)$ is the Fermi distribution function with $f(\varepsilon;\mu,T)=1/\left[1+e^{(\varepsilon-\mu)/k_\mathrm{B}T}\right]$, where $\mu$ is the chemical potential and $M_{i}^{ss'}(\bm{k})=\langle s, \bm{k}|\hbar \hat{v}_{i}| s', \bm{k}\rangle$ is the velocity matrix element with the velocity operator $\hat{v}_{i}$ obtained from the relation $\hat{v}_{i}= \partial \hat{\mathcal{H}} / \partial (\hbar k_{i})$. Here $\Gamma=\hbar/\tau$ is a phenomenological parameter with $\tau$ representing the quasiparticle lifetime limited by disorders or interactions, and the clean noninteracting limit corresponds to $\Gamma\rightarrow0$. 

\section{Zero-Temperature Optical Conductivity} \label{sec:zeroTopcd}

\subsection{Drude Weight}
\begin{figure}
    \centering
    \includegraphics[width=\linewidth]{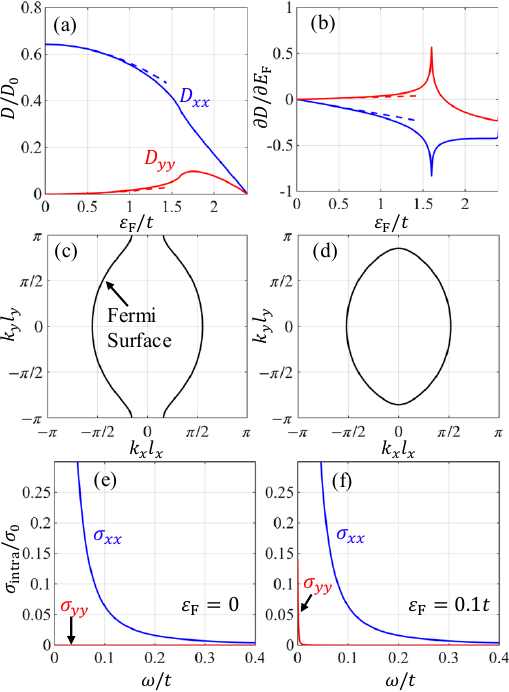}
    \caption{(a) Drude weight as a function of the Fermi energy in units of $D_0=te^2/\hbar^2$, along the $x$ (blue) and $y$ (red) directions. (b) Derivative of the Drude weight with respect to Fermi energy in the units of $D_0/t$, exhibiting a sharp peak at $\EF=1.6t$ where the Lifshitz transition of the Fermi surface occurs. (c), (d) Fermi-surface contours below and above the Lifshitz transition, respectively, showing the change from open to closed topology. (e), (f) Real parts of the intraband optical conductivity along the $x$-(blue) and $y$-(red) directions at charge neutrality ($\EF=0$) and at finite doping ($\EF=0.1t$) with $\Gamma=0.001t$. Here, the optical conductivity is normalized by $\sigma_0=e^2/\hbar$.}
    \label{fig:Drude_weight}
\end{figure}

In this subsection, we examine the intraband contribution to the optical conductivity at zero temperature. For intraband transitions, the term $(f_{s, \bm{k}}-f_{s', \bm{k}})/(\varepsilon_{s, \bm{k}}-\varepsilon_{s', \bm{k}})$ in Eq.~(\ref{eq:KuboFormula}) reduces to the derivative of the Fermi distribution and the velocity matrix element simplifies to $M_{i}^{ss}(\bm{k})= \partial \varepsilon_{s,\bm{k}} / \partial k_i$. Then, the optical conductivity takes the form $\sigma^{\mathrm{intra}}_{ii}(\omega)=i\hbar D_{ii}(\mu)/(\hbar \omega+i\Gamma)$ with the Drude weight 
\begin{equation}
    D_{ii}(\mu)=e^2\int_{-\infty}^{\infty}d\varepsilon\;\Phi_{ii}(\varepsilon)\left[-\frac{\partial f_{s,\bm k}}{\partial\varepsilon}\right].
    \label{eq:Drude_weight_general_formula}
\end{equation}
where 
\begin{equation}
    \Phi_{ii}(\varepsilon)= g
    \sum_s\int\frac{d^2k}{(2\pi)^2}\,v_{s,i}(\bm k)^2\,\delta\!\big(\varepsilon-\varepsilon_{s,\bm k}\big)
\end{equation}
is the transport spectral function.
At zero temperature, the derivative of the Fermi-Dirac distribution function with respect to Fermi energy becomes a negative delta function, i.e., $-\partial f_{s,\bm k}/\partial\varepsilon_{s,\bm k}\rightarrow \delta(\varepsilon_{s,\bm k}-\EF)$. Thus the relation between the Drude weight $D_{ii}(\mu)$ and the transport spectral function $\Phi_{ii}(\varepsilon)$ reduces to 
\begin{equation}
    D_{ii}(\EF)=e^2\Phi_{ii}(\EF),
    \label{eq:drude_spectral}
\end{equation} 
where $\EF$ is the Fermi energy. We restrict the following analysis to $\EF \geq 0$, where only the conduction band ($s = +$) crosses the Fermi level. This restriction follows from the electron-hole symmetry of the Dirac SSH model, which ensures that the optical conductivity depends only on $|\EF|$.

We first consider the low-doping regime $\EF\ll\abs{t-t'}$, where we obtain the Drude weight up to the second order of $\EF$ with $D_{ii}=D^0_{ii}+\delta D_{ii}\EF^2$. We recast $\Phi_{ii}(\EF)$ into the form
\begin{equation} 
    \Phi_{ii}(\EF)=\int_{-\pi/l_y}^{\pi/l_y}\frac{dk_y}{2\pi} \Phi^\mathrm{1D}_{ii}(\EF;k_y)
    \label{eq:Phi}
\end{equation} 
where 
\begin{equation}
    \Phi^\mathrm{1D}_{ii}(\EF;k_y)= g
    \sum_s\int\frac{dk_x}{2\pi}\,v_{s,i}(\bm k)^2\,\delta\!\left(\EF-\varepsilon_{s,\bm k}\right),
    \label{eq:Phi1D}
\end{equation}
is the transport spectral function of an isolated 1D Dirac chain at transverse momentum $k_y$.
In the leading-order linear approximation, the energy dispersion of the conduction band near the nodal line is $\varepsilon_{+,\bm k} \approx \hbar v_\mathrm{F}(k_y)|k_x|$ [see Eq.~(\ref{eq:vF_qy}) for $v_\mathrm{F}(k_y)$], and the $k_x$ integral in Eq.~\eqref{eq:Phi1D} can easily be performed analytically, giving $\Phi^{\mathrm{1D}}_{xx}(k_y) = g\,v_\mathrm{F}(k_y)/(\pi\hbar)$, and the 1D Drude weight
\begin{equation}
    D^{\mathrm{1D}}_{xx}(k_y) = g\frac{e^2 v_\mathrm{F}(k_y)}{\pi\hbar}.
    \label{eq:D1D_xx}
\end{equation}
This expression is precisely the Drude weight of a 1D massless Dirac liquid \cite{Giamarchi2003}, where the appearance of $\hbar$ reflects the quantum nature. Note that the 1D Drude weight is independent of $\EF$ because for 1D Dirac fermions the Fermi velocity is always $\pm v_\mathrm{F}$, regardless of the doping level relative to charge neutrality. 
Using Eqs.~\eqref{eq:drude_spectral} and \eqref{eq:Phi1D} with the obtained $\Phi^{\mathrm{1D}}_{xx}$ immediately yields
\begin{equation}
    D^0_{xx}
    = \frac{2ge^2}{\pi^2\hbar^2}\,
    \frac{l_x}{l_y}\,(t+t')\,E(m),
    \label{eq:Dxx_zero}
\end{equation}
where we use the convention $E(m) = \int_0^{\pi/2}\!\sqrt{1-m\sin^2\theta}\,d\theta$. The elliptic parameter is related to $t$ and $t'$ through
\begin{equation}
    m = \frac{4tt'}{(t+t')^2}.
    \label{eq:modulus}
\end{equation}
Note that Eq.~\eqref{eq:Dxx_zero}, obtained with the linearized dispersion, gives the $\EF$-independent leading term $D^0_{xx}$ in the expansion $D_{xx}(\EF)=D^0_{xx}+\delta D_{xx}\EF^2$.
This is because $D_{xx}$ is simply the $k_y$-average of $D^{\mathrm{1D}}_{xx}(k_y)$, thus inheriting its $\EF$-independence.

The leading $\EF$ correction to $D_{xx}$, i.e., $\delta D_{xx}$, arises from the nonlinearity of the band dispersion away from the nodal line. Expanding the energy dispersion to third order in $k_x$ and using the complete elliptic integral of the first kind $K(m) = \int_0^{\pi/2}\!(1-m\sin^2\theta)^{-1/2}\,d\theta$, we obtain (see Appendix~\ref{sec:Calculation_details_Drude} for calculation details)
\begin{equation}
    \begin{split}
        D_{xx}(\EF) &= \frac{2ge^2}{\pi^2\hbar^2}\frac{l_x}{l_y}
        \left[\left(t+t'\right)\,E(m)
        - \frac{K(m)}{8(t+t')}\EF^2\right],
    \end{split}
    \label{eq:Dxx_full}
\end{equation}
The negative $\EF^2$ correction arises because at finite doping the Fermi velocity is reduced below its linear-dispersion value $v_\mathrm{F}(k_y)$ [Eq.~\eqref{eq:vx_star}].

The Drude weight along the $y$-direction in the low-doping limit can also be obtained asymptotically using the same approach, as shown in Appendix~\ref{sec:Calculation_details_Drude}. The Drude weight along the $y$ direction is given by
\begin{equation}
    D_{yy}(\EF) = \frac{ge^2}{3\hbar^2\pi^2}\frac{l_y}{l_x}
    \frac{(t^2+t^{\prime 2})E(m)-(t-t')^2K(m)}{(t-t')^2(t+t')}\EF^2.
    \label{eq:Dyy_full}
\end{equation}
It should be noted that $D_{yy}$ vanishes at charge neutrality and grows as $\EF^2$ in contrast to $D_{xx}$, which is finite at charge neutrality and decreases as $\EF^2$. Moreover, in the small-$t'$ limit, $t'\ll t$ (corresponding to a large number of layers), $D_{xx}$ approaches $gD_0l_x/\pi l_y $, whereas $D_{yy}\to 0$, where $D_0=te^2/\hbar^2$. This behavior of the Drude weight is a direct consequence of the quasi-1D nature of the band structure in the small-$t'$ limit: at leading order, the system behaves as a collection of 1D Dirac nodes along the $x$ direction, thus suppressing all $y$-directed transport while leaving $x$-directed transport finite.

Thus far, we have focused on the Drude weight in the low-doping regime near the nodal line. In the following, we present our numerical results beyond the low-doping limit obtained via numerical integration. In Fig.~\ref{fig:Drude_weight}(a) we plot the Drude weight calculated over the entire range of the Fermi energy. For small $\EF\ll \abs{t-t'}$, our low-energy asymptotic formulas are in good agreement with the numerical results. It is important to note that at $\EF=2\abs{t-t'}$, the Drude weight exhibits a kink, which is seen more clearly in Fig.~\ref{fig:Drude_weight}(b) where we plot the derivative of the Drude weight with respect to Fermi energy.
This kink appears as a signature of the transition between two distinct Fermi-surface topologies: for $\EF<2\abs{t-t'}$, the Fermi surface is open, whereas for $\EF>2\abs{t-t'}$ it is closed [see Fig.~\ref{fig:Drude_weight}(c) and (d)]. 
It is worth noting that both $D_{xx}$ and $D_{yy}$ decrease with increasing $\varepsilon_{\mathrm F}$ at high Fermi energies. This behavior originates from the lattice nature of the band structure beyond the low-energy linearized regime. As $\varepsilon_{\mathrm F}$ moves away from the nodal line and approaches the upper band edge, the band becomes flatter with the corresponding Fermi velocity reduced. Note that this reduction suppresses both $D_{xx}$ and $D_{yy}$ because the Drude weight is determined by the Fermi-surface average of the squared velocity. 
In Fig.~\ref{fig:Drude_weight}(e) and (f), we present the intraband optical conductivity for the undoped and doped cases, respectively, showing how the features of the Drude weight are manifested in the optical response. Note that for the undoped case, $\sigma_{yy}=0$ over the entire frequency range because $D_{yy}\propto \EF^2$ is zero at charge neutrality, whereas $\sigma_{xx}$ remains finite even at charge neutrality, barely changing with increasing $\EF$, as expected from the analytic Drude-weight results discussed above.

\begin{figure}
    \centering
    \includegraphics[width=\linewidth]{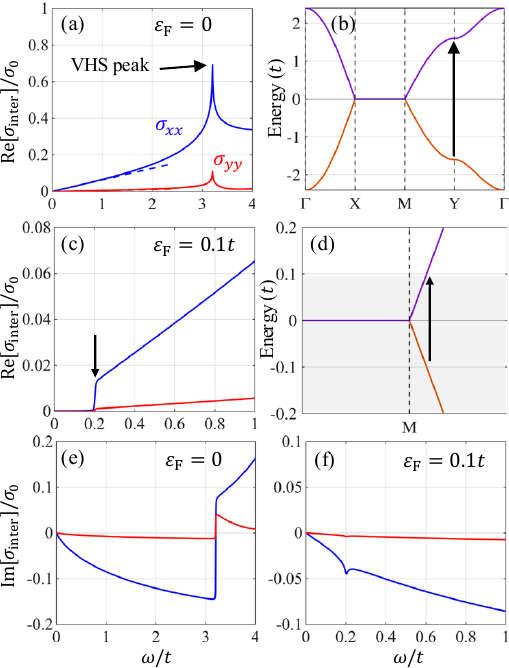}
    \caption{(a) Calculated real parts of the interband optical conductivity at $\EF=0$. A sharp peak appears at frequencies corresponding to interband transitions at van Hove singularity points. The blue (red) solid lines represent the conductivity along the $x$ ($y$) direction. The dashed lines indicate the analytical results at low frequencies given by Eq.~(\ref{eq:inter_sigma_xx}) for $\sigma_{xx}$ and Eq.~(\ref{eq:inter_sigma_yy}) for $\sigma_{yy}$. (b) The energy dispersion with the black arrow indicating transitions at the van Hove singularity points. (c) Calculated real parts optical conductivity at finite doping with $\EF = 0.1t$, exhibiting an optical gap for $\hbar\omega<2\EF$ due to Pauli blocking. (d) Magnified view of the band structure near the $M$ point, where the gray shaded area represents the occupied states. The black arrow denotes interband transitions corresponding to the onset of the optical conductivity at $\hbar\omega=2\EF$. (e) and (f) show calculated imaginary parts of the interband optical conductivity at $\EF = 0$ and $\EF = 0.1t$, respectively. Here, the optical conductivity is normalized by $\sigma_0=e^2/\hbar$.
    } \label{fig:interband_conductivity}
\end{figure}

\subsection{Interband optical conductivity}
We now consider the interband contribution to the optical conductivity. As in the previous section, we first present asymptotic analytic results in the low-frequency regime ($\hbar\omega \ll |t-t'|$), where interband transitions occur predominantly near the nodal line. For the analytical derivation we take the clean limit ($\Gamma\to 0$). Although this approximation would produce a very small quantitative deviation from the numerical results shown in Fig.~\ref{fig:interband_conductivity}, where $\Gamma=0.001t$ is used, it does not alter the qualitative behavior because the relevant photon energies for interband transitions are much larger than $\Gamma$.

Within the clean limit, the real part of the interband optical conductivity follows from the Kubo formula [Eq.~(\ref{eq:KuboFormula})] as
\begin{equation}
    \begin{split}
    \operatorname{Re}\sigma_{ii}^\mathrm{inter}(\omega)
    &= g\frac{\pi e^{2}}{\hbar}
    \sum_{s \neq s'} \int_{\mathrm{BZ}} \frac{d^{2}\bm{k}}{(2\pi)^{2}}\,
    \frac{f_{s,\bm{k}}-f_{s',\bm{k}}}
         {\Delta\varepsilon_{s's}(\bm{k})} \\
    &\quad \times
    \bigl|M_{i}^{ss'}(\bm{k})\bigr|^2\,
    \delta\!\left(\hbar\omega-\Delta\varepsilon_{s's}(\bm{k})\right),
    \end{split}
    \label{eq:interband_Re_sigma}
\end{equation}
$\Delta\varepsilon_{s's}(\bm{k}) = \varepsilon_{s'}(\bm{k}) -
\varepsilon_s(\bm{k})$. Note that on-shell,
$\Delta\varepsilon_{s's} = \hbar\omega > 0$, and thus only the $(s,s')=(-,+)$
pair contributes. For simplicity, we work at charge neutrality ($\EF = 0$). Note that a finite doping suppresses interband absorption for $\hbar\omega < 2\EF$ through Pauli blocking, opening a sharp optical gap at $\hbar\omega = 2\EF$, as shown in our numerical results provided in Fig.~\ref{fig:interband_conductivity}(c).

The real part of the optical conductivity arising solely from interband transitions along the $x$-direction is given by [see Appendix.~\ref{sec:Calculation_details_inter} for the details of derivation] for computation details]
\begin{equation}
    \operatorname{Re}\sigma_{xx}^\mathrm{inter}(\omega)
    = g\frac{e^2}{\hbar}\,\frac{l_x}{l_y}\,
      \frac{\hbar\omega}{16\pi(t+t')}\,K(m).
    \label{eq:inter_sigma_xx}
\end{equation}
and along the $y$-direction, given by
\begin{equation}
    \operatorname{Re}\sigma_{yy}^\mathrm{inter}(\omega)
    = g\frac{e^2}{\hbar}\,\frac{l_y}{l_x}\,
      \frac{\hbar\omega}{24\pi(t+t')}
      \!\left[K(m)
       + \frac{5t^{\prime 2}-t^2}{(t-t')^2}\,E(m)\right].
    \label{eq:inter_sigma_yy}
\end{equation}
It is worth comparing our results with those of other nodal-line semimetals. For a 2D nodal-line semimetal whose nodal ring is protected by $\mathcal{PT}$ symmetry or mirror symmetry, no interband optical transitions can be induced and the optical conductivity is purely contributed from intraband transitions \cite{Barati2017OpticalConductivityNodalLine}. For a 3D nodal-line semimetal, the interband conductivity approaches to a frequency-independent constant in the $\omega\to0$ limit along the axial direction, and increases linearly with $\sigma\propto\omega$ along the radial direction at low frequencies~ \cite{Barati2017OpticalConductivityNodalLine}. In contrast, the interband conductivities along both directions of the present 2D nonsymmorphic nodal-line semimetal, i.e., $\operatorname{Re}\sigma_{xx}$ and $\operatorname{Re}\sigma_{yy}$, grow linearly with $\omega$ at low frequencies [Eqs.~(\ref{eq:inter_sigma_xx}) and~(\ref{eq:inter_sigma_yy})], providing a distinctive optical signature of the nonsymmorphic nodal-line physics that is absent in both 2D symmorphic and 3D nodal-line systems.

Figure~\ref{fig:interband_conductivity}(a) shows our numerically calculated
results at charge neutrality (i.e., $\EF=0$). Note that both conductivities grow linearly at low frequencies, consistent with the analytic formulas derived above (dashed lines). At high frequency, this low-frequency linear behavior is modified by the full lattice band structure. In particular, a sharp peak at higher frequencies arises from transitions near the van Hove singularity, at which the joint density of states diverges [Fig.~\ref{fig:interband_conductivity}(a) and (b)]. 
At finite doping [Fig.~\ref{fig:interband_conductivity}(c), $\EF = 0.1t$], interband transitions are suppressed for $\hbar\omega < 2\EF$ with a sharp onset at $\hbar\omega = 2\EF$, as illustrated in Fig.~\ref{fig:interband_conductivity}(d).

Note that, in contrast to the intraband optical conductivity whose anisotropy appears very strongly with vanishing $D_{yy}$ but $D_{xx}$ remaining finite at charge neutrality, the interband optical conductivity is much less anisotropic than the intraband optical conductivity, growing linearly with $\omega$ in the low-frequency regime for both $\sigma_{xx}$ and $\sigma_{yy}$. This is because the interband optical conductivity is determined by the joint density of states and the interband velocity matrix elements given in Eqs.~(\ref{eq:Mex}) and~(\ref{eq:Mey}), rather than by the highly direction-dependent Fermi velocity that determines the intraband optical conductivity. Note that the interband velocity matrix elements are generally direction dependent. However, for the system considered here, both $|M_x^{-+}(\bm{k})|^2$ and $|M_y^{-+}(\bm{k})|^2$ contain the same factor $\varepsilon_{+,\bm{k}}^2$, which, under the optical transition condition $\varepsilon_{+,\bm{k}}=\hbar\omega/2$, leads to the same linear low-frequency dependence.

\section{Finite-temperature effects} \label{sec:finiteTopcd}
In this section, we extend our previous analysis of the zero-temperature results to finite temperature $T>0$. At finite temperature, the Drude weight can be calculated using Eq.~(\ref{eq:Drude_weight_general_formula}) with the chemical potential $\mu(T)$ determined from the condition of fixed carrier density measured relative to charge neutrality:
\begin{equation}
    n_\mathrm{ex}=
    \int_{0}^{\infty} d\varepsilon\,\mathcal{D}(\varepsilon)
    \left[
        f(\varepsilon;\mu,T)-f(\varepsilon;-\mu,T)
    \right].
    \label{eq:n_condition}
\end{equation}
Here, $n_\mathrm{ex}$ denotes the excess carrier density relative to charge neutrality.
Within the linear approximation $\varepsilon_{\pm,\bm k}\approx \pm \hbar v_\mathrm{F}(k_y)|k_x|$, the low-energy density of states is constant,
\begin{equation} 
    \mathcal{D}(\varepsilon)=\mathcal{D}_0
    =\frac{2gK(m)}{\pi^2 l_x l_y (t+t')}.
    \label{eq:DOS_linear}
\end{equation}
Applying the Sommerfeld expansion to Eq.~(\ref{eq:n_condition}), we obtain
\begin{equation} \label{eq:sommerfeld_n}
    n_\mathrm{ex} = \mathcal{D}_0\mu(T) + \frac{\pi^2}{6}(k_\mathrm{B}T)^2
    \frac{\partial \mathcal{D}(\varepsilon)}{\partial \varepsilon}\bigg|_{\varepsilon=\mu(T)} + \mathcal{O}(T^4)
\end{equation} 
With $\partial \mathcal{D}(\varepsilon)/\partial \varepsilon=0$ for the constant low-energy density of states, $n_\mathrm{ex}= \mathcal{D}_0\mu(T)$.
Since $n_\mathrm{ex} = \mathcal{D}_0\EF$ at $T=0$, we obtain 
\begin{equation}
    \mu(T) = \EF,
    \label{eq:mu_linear}
\end{equation}
indicating that the chemical potential does not shift with temperature at any order in $T$ within the linear approximation. 
This extremely weak temperature dependence of the chemical potential comes from the constant density of states [Eq.~(\ref{eq:DOS_linear})], and is similar to that of a conventional 2D electron gas with parabolic dispersion, where the density of states is also constant. 
This is qualitatively distinct from systems with an energy-dependent density of states, such as graphene and a 3D electron gas, where $\mu$ shifts at leading order as $\mu \approx \EF[1-\pi^2(k_\mathrm{B}T)^2/6\EF^2]$ and $\mu \approx \EF[1 - \pi^2(k_\mathrm{B}T)^2/12\EF^2]$, respectively.

Including the cubic term in the dispersion [Eq.~(\ref{eq:cubic_disp})],
the density of states acquires an energy-dependent correction. Performing the $k_x$ integration with the cubic dispersion, retaining terms up to the second order of $\varepsilon$, gives
\begin{equation}
    \mathcal{D}(\varepsilon) \approx \mathcal{D}_0
    \left(1 + \beta\varepsilon^2\right),
    \label{eq:DOS_cubic}
\end{equation}
where
\begin{equation}
    \beta = \frac{E(m)}{8K(m)(t-t')^2}.
    \label{eq:beta}
\end{equation}
Substituting Eq.~(\ref{eq:DOS_cubic}) into the Sommerfeld expansion of Eq.~(\ref{eq:n_condition}), and using the fixed-density condition, we obtain the chemical potential to order $T^2$ after straightforward algebra:
\begin{equation}
    \mu(T) \approx \EF
    - \frac{\pi^2\beta}{3}\,
    \frac{\EF(k_\mathrm{B}T)^2}{1+\beta\EF^2}.
    \label{eq:mu_cubic}
\end{equation}
Note that the correction is of relative order $\beta\EF^2 \sim \EF^2/(t-t')^2$, which is small in the low-doping regime $\EF \ll |t-t'|$.


We now calculate the finite-temperature Drude weight including the
shift of the chemical potential derived in Eq.~(\ref{eq:mu_cubic}).
Using the Sommerfeld expansion, we approximate the Drude weight at finite temperature as
\begin{equation}
    D_{ii}(T) \approx D_{ii}(\mu)
    + \frac{\pi^2}{6}(k_\mathrm{B}T)^2 D''_{ii}(\mu)
    \label{eq:Sommerfeld_D}
\end{equation}
where primes denote differentiation with respect to $\mu$. By expanding $D_{ii}(\mu) = D_{ii}(\EF) + D'_{ii}(\EF)\delta\mu + \mathcal{O}(\delta\mu^2)$ with $\delta\mu = \mu(T) - \EF$ given by Eq.~\eqref{eq:mu_cubic}, we obtain $D_{xx}(\mu) = D_{xx}^0 - C_{xx}\mu^2$, where
\begin{equation}
    D_{xx}^0 = \frac{2ge^2}{\pi^2\hbar^2}\frac{l_x}{l_y}(t+t')E(m),
\end{equation}
and
\begin{equation}
    C_{xx} = \frac{ge^2l_x}{\hbar^2l_y}
    \frac{K(m)}{4\pi^2(t+t')}.
    \label{eq:Cxx_def}
\end{equation}
Substituting $D''_{xx}(\mu) = -2C_{xx}$ into Eq.~\eqref{eq:Sommerfeld_D}, we obtain the finite-temperature Drude weight along the $x$ direction,
\begin{widetext}
\begin{equation}
    D_{xx}(T) = \frac{2ge^2}{\pi^2\hbar^2}\frac{l_x}{l_y}(t+t')E(m)
    - \frac{ge^2l_x K(m)}{4\pi^2\hbar^2l_y(t+t')}
    \left[\EF^2
    + \frac{\pi^2(k_\mathrm{B}T)^2}{3}\,
    \frac{1-\beta\EF^2}{1+\beta\EF^2}
    \right].
    \label{eq:Dxx_T}
\end{equation}
\end{widetext}

Using a similar algebraic calculation, we obtain $D_{yy}(\mu) = C_{yy}\mu^2$, where
\begin{equation}
    C_{yy} = \frac{ge^2l_y}{\hbar^2l_x}
    \frac{(t^2+t^{\prime 2})E(m)-(t-t')^2K(m)}
         {3\pi^2(t-t')^2(t+t')},
    \label{eq:Cyy_def}
\end{equation}
With $D''_{yy} = 2C_{yy}$ substituted into Eq.~\eqref{eq:Sommerfeld_D}, the finite-temperature Drude weight along the $y$ direction is then
\begin{widetext}
    \begin{equation}
        D_{yy}(T) = \frac{ge^2l_y}{\hbar^2l_x}
        \frac{(t^2+t^{\prime 2})E(m)-(t-t')^2K(m)}
             {3\pi^2(t-t')^2(t+t')}
        \left[\EF^2
        + \frac{\pi^2(k_\mathrm{B}T)^2}{3}\,
        \frac{1-\beta\EF^2}{1+\beta\EF^2}
        \right].
        \label{eq:Dyy_T}
    \end{equation}    
\end{widetext}
It is worth noting that the finite-temperature effects are strongly anisotropic and opposite in sign: $D_{xx}$ decreases with $T$, while $D_{yy}$ increases with $T$.
To order $T^2$, both results for $D_{xx}$ and $D_{yy}$ can be obtained from the zero-temperature Drude weights by the substitution
\begin{equation}
    \EF^2 \to
    \EF^2 + \frac{\pi^2(k_\mathrm{B}T)^2}{3}\,
    \frac{1-\beta\EF^2}{1+\beta\EF^2}.
    \label{eq:Sommerfeld_sub}
\end{equation}
In the low-doping limit where the linear band approximation is valid, the substitution reduces to the simple replacement $\EF^2\to\EF^2 + \pi^2(k_\mathrm{B}T)^2/3$, which is the standard Sommerfeld result for a system with an energy-independent density of states. The correction factor $(1-\beta\EF^2)/(1+\beta\EF^2)$ captures the suppression of the thermal correction due to the downward shift $\delta\mu < 0$ [Eq.~(\ref{eq:mu_cubic})] arising from the nonlinear band curvature beyond the linear band regime.

For the experimentally estimated parameters of the $n=2$ member of the \matname family, $t\sim0.177~\mathrm{eV}$ and $t'\sim0.05~\mathrm{eV}$ \cite{zhangObservationDimensioncrossoverTunable2022}, we obtain $\beta\simeq4.71~\mathrm{eV}^{-2}$. This value yields a negligibly small chemical-potential shift at room temperature, with $|\delta\mu|/|\EF|\lesssim10^{-2}$ and $|\delta\mu|/(k_{\mathrm B}T)<0.1$. As one can see straightforwardly from Eq.~(\ref{eq:beta}), $\beta$ decreases with decreasing $t'$ (and thus increasing $n$), thus indicating that the chemical-potential shift is expected to remain small for other members of the \matname family as well.
This suggests that the zero-temperature results of Eqs.~(\ref{eq:Dxx_full}) and~(\ref{eq:Dyy_full}) provide an excellent approximation at all experimentally accessible temperatures.

In the following, we discuss the finite-temperature effects on the interband optical conductivity. At finite temperature, evaluating the difference between the Fermi occupation factors, $f_{-,\mathbf{k}}-f_{+,\mathbf{k}}$, in Eq.~(\ref{eq:interband_Re_sigma}) under the optical transition condition $\hbar\omega=2\varepsilon_{+,\bm{k}}$ gives
\begin{equation} \label{eq:thermal_factor}
    \mathcal{F}(\omega,\mu,T)
    \equiv
    \frac{
    \sinh\left(\hbar\omega/2k_{\mathrm B}T\right)
    }{
    \cosh\left(\mu/k_{\mathrm B}T\right)
    +
    \cosh\left(\hbar\omega/2k_{\mathrm B}T\right)
    },
\end{equation}
with $\mu$ given by Eq.~(\ref{eq:mu_cubic}).
By substituting this into the Kubo formula and following the derivation of the zero-temperature results, we obtain the finite-temperature optical conductivities as
\begin{equation}
    \operatorname{Re}\sigma_{xx}(\omega,T) =
    g\frac{e^2}{\hbar}\frac{l_x}{l_y}
    \frac{\hbar\omega}{16\pi(t+t')}K(m)
    \mathcal{F}(\omega,\mu,T),
\end{equation}
and
\begin{align}
    \operatorname{Re}\sigma_{yy}(\omega,T) = {} 
    & g\frac{e^2}{\hbar}\frac{l_y}{l_x} \frac{\hbar\omega}{24\pi(t+t')} \nonumber \\
    & \times \left[ K(m) + \frac{5t^{\prime 2}-t^2}{(t-t')^2}E(m) \right] \mathcal{F}(\omega,\mu,T).
\end{align}
This leads to two distinct finite-temperature effects. First, the sharp optical gap at $\hbar\omega = 2\EF$ is replaced by a thermally smeared edge at $\hbar\omega = 2\mu(T)$
with the gap edge shifting slightly downward from $2\EF$ to $2\mu(T) < 2\EF$ at finite $T$, by an amount $2|\delta\mu|$ [Eq.~(\ref{eq:mu_cubic})]. Second, the low-frequency analytic results [Eqs.~(\ref{eq:inter_sigma_xx}) and~(\ref{eq:inter_sigma_yy})], which were derived at $\EF = 0$, are recovered for $\hbar\omega\gg k_\mathrm{B}T$ because the thermal factor[Eq.~(\ref{eq:thermal_factor})] approaches to unity. 
For $\hbar\omega\ll k_\mathrm{B}T$ at charge neutrality (i.e., $\mu=0$), the thermal factor in Eq.~(\ref{eq:thermal_factor}) reduces to $\hbar\omega/4k_\mathrm{B}T$. Since the zero-temperature interband conductivity is linear in $\omega$, this implies a crossover from $\sigma\propto\omega$ for $\hbar\omega\gg k_\mathrm{B}T$ to $\sigma\propto\omega^2/T$ for $\hbar\omega\ll k_\mathrm{B}T$. Such a crossover may be observable in sufficiently clean samples with $\Gamma\ll\hbar\omega\ll k_\mathrm{B}T$, where the interband optical conductivity is not entirely obscured by the Drude peak. In practice, the crossover should be easier to observe in the $y$-direction response, where the Drude peak remains much weaker than in the $x$ direction, as already seen in the zero-temperature results.

\section{Discussion and Conclusion} \label{sec:conclusion}

Before concluding, we make two remarks. First, the nodal-line state of the \matname family is associated with a nontrivial Berry phase for a closed loop linking the nodal line with the line crossing itself protected by the nonsymmorphic glide-mirror symmetry. Note that geometric quantities such as Berry phase usually do not appear as a separate quantity in the longitudinal linear optical responses, as shown in our results where the low-energy interband response and the anisotropic Drude weight are not directly related to the topological invariant. However, the optical conductivity calculated in this work indirectly probes the consequences of the symmetry-protected gapless band structure, such as the absence of a finite optical gap at charge neutrality.

Second, our results for the Drude weight are closely related to the plasmon dispersion studied in recent works. This relation follows from the continuity equation, which directly connects the density-density polarization function to the longitudinal optical conductivity \cite{mahanManyparticlePhysics2000}:
\begin{equation}\label{eq:continuity}
    \sigma_\mathrm{L}({\bm q},\omega)=
    \frac{i e^2\omega}{q^2}\Pi({\bm q},\omega),
\end{equation}
where $\Pi({\bm q},\omega)$ is the density-density polarization function. 
In the Drude limit, the longitudinal optical conductivity is written as
\begin{equation}\label{eq:opcd_drude}
    \sigma_\mathrm{L}(\omega)
    \simeq
    \frac{iD_\mathrm{L}}{\omega},
\end{equation}
where $D_\mathrm{L}$ is the longitudinal Drude weight. For an anisotropic system, the longitudinal optical conductivity is the projection of the conductivity tensor along the momentum direction, and thus, in the long wavelength limit($q\to0$), is expressed as
\begin{equation}\label{eq:anisotropic_opcd}
\sigma_\mathrm{L}(\omega)
=
\sigma_{xx}(\omega)\cos^2\theta
+
\sigma_{yy}(\omega)\sin^2\theta,
\end{equation}
where ${\bm q}=q(\cos\theta,\sin\theta)$, and the Drude weight correspondingly as
\begin{equation}
D_\mathrm{L}(\theta)
=
D_{xx}\cos^2\theta
+
D_{yy}\sin^2\theta.
\end{equation}
Combining Eqs.~(\ref{eq:continuity})-(\ref{eq:anisotropic_opcd}), one can easily see that the polarization function in the long-wavelength limit is proportional to $D_\mathrm{L} q^2/\omega^2$, leading to the long-wavelength plasmon dispersion satisfying $\omega_p(q,\theta)\propto \sqrt{qD_\mathrm{L}(\theta)}$, which is obtained from the zeros of the dielectric function [$1-V(q)\Pi({\bf q},\omega)=0$] \cite{mahanManyparticlePhysics2000}. This explains the anisotropic plasmon behavior reported in Ref.~\cite{caoPlasmonsTwodimensionalNonsymmorphic2023}: the plasmon dispersion along the direction normal to the nodal-line ($\theta=0$) is weakly dependent on the Fermi energy (i.e., carrier density) with $D_{xx}$ independent of $\varepsilon_\mathrm{F}$ in the low-doping regime, while along the nodal-line direction, the plasmon dispersion varies depending on the Fermi energy with $D_{yy}\sim \varepsilon_{F}^2$.

In this work, we have studied the optical conductivity of the \matname family materials.
We first investigated the intraband optical conductivity. At low doping  $\EF\ll |t-t'|$, we show that the Drude weight normal to the nodal line is finite at charge neutrality with Planck's constant appearing, implying that the longitudinal Drude weight inherits the quantum nature of 1D Dirac physics. In contrast, the longitudinal Drude weight vanishes quadratically with Fermi energy, as in an ordinary metal. In the high-doping regime, we showed that a kink structure in the Drude weight arises from the Lifshitz transition between open and closed Fermi surfaces.
For the interband optical conductivity, we showed that both $\operatorname{Re}\sigma_{xx}$ and $\operatorname{Re}\sigma_{yy}$ shows linear frequency dependence in the low-frequency regime with only a direction-dependent slope, despite the anisotropic quasi-1D electronic structure.  At higher frequencies, both conductivities develop a sharp peak arising from interband transitions near the van Hove singularity.

Regarding finite-temperature effects, we showed that, within the linear band approximation, the chemical potential does not deviate from the Fermi energy due to the energy-independent density of states. Accounting for the nonlinear band curvature (up to cubic order), however, we showed that the chemical potential shifts downward and the resulting finite-temperature corrections to the Drude weight can be captured by the simple substitution $\EF^2 \to \EF^2 + \frac{\pi^2(k_\mathrm{B}T)^2}{3}\frac{1-\beta\EF^2}{1+\beta\EF^2}$ in the zero-temperature result. For typical material parameters of the \matname family, the thermal corrections at room temperature are several orders of magnitude smaller than $t^2$, and thus the zero-temperature results should provide an excellent approximation at experimentally accessible temperatures.

In conclusion, we reveal optical fingerprints of the 1D nature of the \matname family of nodal-line semimetals, providing both physical insight into their optical response and concrete guidance for their experimental identification.

\appendix
\section{Calculation of the Drude weight} \label{sec:Calculation_details_Drude}
\subsection{\texorpdfstring{$D_{xx}(\EF)$}{Dxx(EF)}}
To obtain the Drude weight to second order of $\EF$, we first expand the energy dispersion up to the third order in $k_x$: 
\begin{equation}
    \varepsilon_{\pm,\bm{k}} \approx \pm\hbar v_\mathrm{F}(k_y)|k_x|
    \mp \frac{\hbar v_\mathrm{F}(k_y)l_x^2}{24}\,|k_x|^3,
    \label{eq:cubic_disp}
\end{equation}
which shifts the Fermi momentum and thereby modifies the Fermi velocity. 
The leading-order solution of $\varepsilon_{+,\bm{k}} = \EF$ in Eq.~\eqref{eq:cubic_disp}  for $k_x$ is $k_x^*\approx k_0 + k_0^3 l_x^2/24$ where $k_0= \EF/\hbar v_\mathrm{F}(k_y)$. The Fermi velocity up to the second order of $\EF$ is written as
\begin{equation}
    \begin{split}
        v_x(k_x^{*}, k_y)
        &= \frac{1}{\hbar}\left.\frac{\partial\varepsilon_{{+},\bm k}}{\partial k_x}\right|_{k_x^*}\\
        &\approx v_\mathrm{F}(k_y)\!\left[1 - \frac{l_x^2\EF^2}{8\hbar^2 v_\mathrm{F}^2(k_y)}\right].
        \label{eq:vx_star}
    \end{split}
\end{equation}
Evaluating the 1D delta-function integral in $\Phi_{xx}^{\mathrm{1D}}$ [Eq.~\eqref{eq:Phi1D}] with the two symmetric Fermi-surface roots $\pm k_x^{*}$, we obtain the quadratic correction in $\EF$ to the transport spectral function
\begin{equation}
    \delta\Phi^{\mathrm{1D}}_{xx}(\EF;k_y)
    = -\frac{gl_x}{\pi\hbar^2}\frac{\EF^2}{8\Abk}.
    \label{eq:Phi1D_cubic}
\end{equation}
Then, the Drude weight correction of the second order of $\EF$ is obtained by taking the $k_y$-average of $\delta\Phi^{\mathrm{1D}}_{xx}(\EF;k_y)$ and Eq.~(\ref{eq:Drude_weight_general_formula}):
\begin{equation}
    \delta D_{xx}(\EF) = -\frac{ge^2l_x}{\pi\hbar^2}
             \int_{-\pi/l_y}^{\pi/l_y}\!\!\frac{dk_y}{2\pi}
              \frac{\EF^2}{8\Abk}.
    \label{eq:Dxx_ky}
\end{equation}
Using the complete elliptic integrals of the first kind $K(m) = \int_0^{\pi/2}\!(1-m\sin^2\theta)^{-1/2}\,d\theta$, $D_{xx}(\EF)$ is expressed as
\begin{equation}
    \begin{split}
        D_{xx}(\EF) &= \frac{2ge^2}{\pi^2\hbar^2}\frac{l_x}{l_y}
        \left[\left(t+t'\right)\,E(m)
        - \frac{K(m)}{8(t+t')}\EF^2\right],
    \end{split}
\end{equation}

\subsection{\texorpdfstring{$D_{yy}(\EF)$}{Dyy(EF)}}
The Drude weight along the $y$-direction can be obtained similarly.
Within the linear band approximation, i.e., $\varepsilon_{\pm,\bm{k}} \approx \pm\hbar v_\mathrm{F}(k_y)|k_x|$, we obtain
\begin{equation}\label{eq:phi_1D_yy}
    \Phi^{\mathrm{1D}}_{yy}(\EF;k_y) = \frac{g(tt')^2 l_y^2 \sin^2(k_y l_y)\, \varepsilon_{\mathrm{F}}^2}{\pi \hbar^2 l_x A(k_y l_y)^5}.
\end{equation}
In order to evaluate the Drude weight through $D_{yy}(\EF) = e^2\Phi_{yy}(\EF)$, we need to take the $k_y$-average of $e^2\Phi^{\mathrm{1D}}_{yy}(\EF)$, which requires calculating
\begin{equation}
    \frac{1}{2\pi l_y}\!\int_{-\pi}^{\pi}
    \frac{\sin^2 \varphi}{A(\varphi)^5}\,d\varphi
    = \frac{8}{\pi l_y(t+t')^5}
      I(m),
    \label{eq:Jintegral}
\end{equation}
where
\begin{equation}
    \begin{split}
        I(m)=&\int_0^{\pi/2}\!\frac{\sin^2\varphi\cos^2\varphi}
                             {(1-m\sin^2\varphi)^{5/2}}\,d\varphi \\
        =& \frac{(2-m)E(m)-2(1-m)K(m)}{3m^2(1-m)}.
    \end{split}
    \label{eq:Isc}
\end{equation}
Substituting Eqs.~(\ref{eq:Jintegral}) and~(\ref{eq:Isc}) into the $k_y$-average of $\Phi^{\mathrm{1D}}_{yy}(k_y;\EF)$, we obtain the Drude weight along the $y$ direction
\begin{equation}
    D_{yy}(\EF) = \frac{ge^2}{3\hbar^2\pi^2}\frac{l_y}{l_x}
    \frac{(t^2+t^{\prime 2})E(m)-(t-t')^2K(m)}{(t-t')^2(t+t')}\EF^2.
\end{equation}

\section{Interband optical conductivity} \label{sec:Calculation_details_inter}
For the $x$- and $y$-components of the matrix elements, straightforward algebra gives
\begin{equation}
    \bigl|M_x^{-+}(\bm{k})\bigr|^2
    = \frac{l_x^2}{4}\,\varepsilon_{+,\bm{k}}^2,
    \label{eq:Mex}
\end{equation}
and
\begin{equation}
    \bigl|M_y^{-+}(\bm{k})\bigr|^2
    = \frac{t^{\prime 2}l_y^2\,\bigl(t'+t\cos(k_yl_y)\bigr)^2}
           {\Abk^4}\,\varepsilon_{+,\bm{k}}^2.
    \label{eq:Mey}
\end{equation}
Both matrix elements are proportional to $\varepsilon_{+,\bm{k}}^2$, differing only in the direction-dependent prefactor. Since $\varepsilon_{+,\bm{k}}^2$ becomes the constant $(\hbar\omega/2)^2$ under the on-shell condition $\Delta\varepsilon_{s's}(\bm{k}) = \hbar\omega$, the momentum dependence is entirely contained in the prefactor. Then, one encounters the following formula during the $k_x$ integration of Eq.~\eqref{eq:interband_Re_sigma}:
\begin{equation}
    \int_{-\Lambda}^{\Lambda}\!\!dk_x\,
    \delta\!\bigl(\hbar\omega - 2|k_x|l_x \Abk\bigr)
    = \frac{1}{l_x \Abk}\,\Theta\!\bigl(\Lambda - k_x^*\bigr) 
    \label{eq:delta_identity}
\end{equation}
where $\Theta$ is the Heaviside function, $k_x^* = \frac{\hbar\omega}{2l_x \Abk}$, and $\Lambda$ is the momentum cutoff for the low-energy effective model. In the low-frequency regime we are considering here, $k_x^* \ll \Lambda$ and thus we drop the step function in the derivations that follow. Substituting Eqs.~(\ref{eq:Mex}) and~(\ref{eq:delta_identity}) into Eq.~(\ref{eq:interband_Re_sigma}) with $\varepsilon_{+,\bm k} = \hbar\omega/2$ and performing the $k_y$ integral via the substitution $\varphi = k_yl_y$, together with the integral result
\begin{equation}
    \int_{-\pi}^{\pi}\!d\varphi\frac{1}{A(\varphi)}
    = \frac{4K(m)}{t+t'},
    \label{eq:Jdef}
\end{equation}
we obtain the real part of the optical conductivity contributed solely by interband transitions along the $x$ axis, given by
\begin{equation}
    \operatorname{Re}\sigma_{xx}(\omega)
    = g\frac{e^2}{\hbar}\,\frac{l_x}{l_y}\,
      \frac{\hbar\omega}{16\pi(t+t')}\,K(m).
\end{equation}
The optical conductivity along the $y$ direction can be obtained similarly. Substituting Eq.~(\ref{eq:Mey}) and the $k_x$ integration result [Eq.~(\ref{eq:delta_identity})] into Eq.~(\ref{eq:interband_Re_sigma}), $\operatorname{Re}\sigma_{yy}(\omega)$ is written as
\begin{equation} \label{eq:sigmayy_Iy}
    \operatorname{Re}\sigma_{yy}(\omega)
    = \frac{ge^2 t^{\prime 2}l_y\omega}{16\pi l_x}
      \,I_y(t,t'),
\end{equation}
where 
\begin{equation}
    I_y(t,t') \equiv \int_{-\pi}^{\pi}\!d\theta\,
    \frac{\bigl(t'+t\cos\theta\bigr)^2}{A(\theta)^5}.
    \label{eq:Iy}
\end{equation}
To evaluate $I_y$ in closed form, we differentiate both sides of Eq.~(\ref{eq:Jdef}) twice with respect to $t'$. Making use of the relation $\partial^2 A^{-1}/\partial t^{\prime 2} = 3(t'+t\cos\theta)^2 A^{-5} - A^{-3}$, we obtain the following equality
\begin{equation}
\begin{split}
    \frac{\partial^2 }{\partial t^{\prime 2}}\left[\frac{4K(m)}{t+t'}\right]
    =& \frac{2}{(t+t')t^{\prime 2}}
      \!\left[K(m) + \frac{3t^{\prime 2}-t^2}{(t-t')^2}\,E(m)\right].\\
    =& 3 I_y(t,t') -  \frac{4}{(t+t')(t-t')^2}\,E(m).
    \end{split}
    \label{eq:Jpp}
\end{equation}
Solving Eq.~\eqref{eq:Jpp} for $I_y$ and substituting it into Eq.~\eqref{eq:sigmayy_Iy}, we obtain
\begin{equation}
    \operatorname{Re}\sigma_{yy}(\omega)
    = g\frac{e^2}{\hbar}\,\frac{l_y}{l_x}\,
      \frac{\hbar\omega}{24\pi(t+t')}
      \!\left[K(m)
       + \frac{5t^{\prime 2}-t^2}{(t-t')^2}\,E(m)\right].
\end{equation}

\acknowledgements
Please replace the entire Acknowledgments section with the following paragraph:
This work was partly supported by the Institute of Information \& Communications Technology Planning \& Evaluation (IITP)-ITRC (Information Technology Research Center) grant funded by the Korea government (MSIT) (IITP-2026-RS-2024-00437284, 20\%) and the National Research Foundation of Korea (NRF) grant funded by the Korea Government (MSIT) (No. RS-2023-00272513, 60\%).
\bibliography{reference_revised}

\end{document}